\documentstyle[12pt]{article}

\begin{document}

\title{Quasi-exact Solvability of the Pauli Equation}

\author{Choon-Lin Ho$^1$ and Pinaki Roy$^2$}
\date
{\small \sl $^1$Department of Physics, Tamkang University, Tamsui
25137, Taiwan\\ $^2$Physics and Applied Mathematics Unit, Indian
Statistical Institute,
Calcutta 700035,
India}

\maketitle

\begin{abstract}
We present a general procedure for determining possible
(nonuniform) magnetic fields such that the Pauli equation becomes
quasi-exactly solvable (QES) with an underlying $sl(2)$ symmetry. This
procedure makes full use of the close connection between QES
systems and supersymmetry.  Of the ten classes of $sl(2)$-based
one-dimensional QES systems, we have found that nine classes allow
such construction.
\end{abstract}

 \vskip 3cm 

\newpage

\section{Introduction}
The Pauli equation describes the motion of a charged particle in
an external magnetic field.  It is given by the Hamiltonian
($\hbar=e=2m_e=c=1$)
\begin{eqnarray}
H=(p_x +A_x)^2 + (p_y + A_y)^2 +\frac{g}{2}(\nabla\times {\bf
A})_z\sigma_z~, \label{Pauli}
\end{eqnarray}
where $p_x,~p_y$ are the momentum operators, $g=2$ is the
gyromagnetic ratio, ${\bf A}$ is the vector potential of the
electromagnetic field, and $\sigma_z$ is the Pauli matrix.  For
uniform magnetic field $B_x=B_y=0,~B_z=B$ the system is exactly
solvable, giving the Landau levels. On the other hand, it was
proved in \cite{AC} that for any general magnetic field
$B_z=B(x,y)$ perpendicular to the $xy$ plane, the ground state is
exactly calculable, owing to the existence of supersymmetry (SUSY)
in eq.(\ref{Pauli}) \cite{Cooper}.  The general result of
\cite{AC} can be viewed as a very special case of the newly
discovered quasi-exactly solvable (QES) systems, which are systems
that allow parts of their spectrum to be solved algebraically
(\cite{TU}-\cite{Ho}).

The Landau problem and the result of \cite{AC} represent the two
extremes of the spectral problem of eq.(\ref{Pauli}). It is thus
of interest to determine if other possibilities exist. Based on
SUSY of the Pauli equation and the idea of shape invariance, it
was shown that there exist three other forms of (nonuniform)
magnetic field which make Pauli equation exactly solvable
\cite{Cooper} (see also \cite{AI}). It seems difficult, if
not impossible, to find other forms of the magnetic field such
that the Pauli equation could be exactly solved.  A more modest
aim is to determine magnetic fields such that parts of the
spectrum of the Pauli equation can be algebraically obtained.  In
other words, one looks for those magnetic fields under which the
Pauli equation becomes quasi-exactly solvable (we note that for
certain non uniform magnetic fields, the Pauli equation admits two
solutions \cite{tka} and thus can be considered as a QES system).
But even with this modest aim, the possibility seems enormous,
since there are many QES systems based on different Lie algebras.
In this paper we would like to make an attempt in this direction
based on the simplest Lie algebra, namely, the $sl(2)$ algebra.
QES systems based on $sl(2)$ algebra have been completely
classified by Turbiner \cite{Tur2}, and the necessary and
sufficient conditions for the normalizability of the wave
functions in such systems were completely determined in
\cite{GKO}. It turns out that general forms of the magnetic field
can be found so that the Pauli equation can be fitted into nine of
the ten classes in \cite{Tur2}.  The magnetic fields giving rise
to these nine classes of QES potentials are divided into two
groups: six in asymmetric gauge, and the other three in
symmetric gauge. We shall describe these cases separately in Sect.
2 and 3.  We would like to mention here that it is not necessary to
consider
magnetic fields that give rise to QES Pauli Hamiltonians with
periodic potentials, since it has been proved in \cite{GKO} that
the wave functions in such systems are not normalizable.

\section{Magnetic field in asymmetric gauge}

Consider magnetic field in the asymmetric gauge given by the
vector potential
\begin{eqnarray}
A_x(x,y)=0~, A_y(x,y)=-{\bar W}(x)~,
\end{eqnarray}
where ${\bar W}(x)$ is an arbitrary function of $x$.  The magnetic field

$\bf B$ has components $B_x=B_y=0$ and $B_z=B(x)=-{\bar W}^\prime (x)$.
The
Pauli Hamiltonian is then given by
\begin{eqnarray}
H=p_x^2 + (p_y-\bar{W}(x))^2 -\bar{W}^\prime(x)\sigma_z~,
\end{eqnarray}
where $\bar{W}^\prime(x)=d\bar{W}(x)/dx$. The eigenfunction
$\tilde \psi$ can be factorized as
\begin{eqnarray}
{\tilde\psi} (x,y)=e^{-iky}{\bar\psi}(x)~.
\end{eqnarray}
Here $k$ ($-\infty <k<\infty$) are the eigenvalues of $p_y$, and
${\bar\psi}(x)$ is a two-component function of $x$. The upper and
lower components of $\bar\psi$ are then governed by the
Hamiltonians $H_-$ and $H_+$ respectively, where
\begin{eqnarray}
H_\mp = -\frac{d^2}{dx^2} +\left(\bar{W}(x) + k\right)^2
\mp\bar{W}^\prime(x)~.
\end{eqnarray}
In this form the SUSY structure of the Pauli equation is clearly
exhibited, with $W(x)=\bar{W}(x)+k$ playing the r\^ole of the
superpotential. Once the upper component $\psi$ of $\bar\psi$ is
solved for a nonzero energy, the lower component can be obtained
by applying appropriate supercharge on the upper component
\cite{Cooper}  and vice versa. Hence, it is suffice to consider
the solution of the upper component, which satisfies the
Schr\"odinger equation (we assume that the upper component has a
normalizable zero energy state) $H_-\psi=E\psi$, where
\begin{eqnarray}
H_- =-\frac{d^2}{dx^2} + V(x) ~,\label{SE}
\end{eqnarray}
with
\begin{eqnarray}
V(x)=W(x)^2 -W^\prime(x)~. \label{V}
\end{eqnarray}
From the knowledge of shape invariant SUSY potentials, it was
found that there are four allowed forms of shape invariant
$\bar{W} (x)$ for which the spectrum of the Pauli equation can be
algebraically written down \cite{Cooper}. One of the four forms
gives rise to uniform magnetic field.

The SUSY structure of the Pauli equation can be made use of in a
different way, namely, in its close connection with quasi-exactly
solvability \cite{Shifman,Roy}.  We shall give a brief description
of this connection below.

Consider a system described by eq.(\ref{SE}).  We shall look for $V(x)$
such
that the system is QES.  According to the theory of QES models, one
first makes
an imaginary gauge transformation on the wave function $\psi (x)$
\begin{eqnarray}
\psi (x)= \phi(x) e^{-g(x)}~,
\end{eqnarray}
where $g(x)$ is the gauge function.  The function $\phi(x)$
satisfies
\begin{eqnarray}
-\frac{d^2\phi(x)}{dx^2} + 2 g^\prime \frac{d\phi(x)}{dx} +
\left[V(x)+ g^{\prime\prime} - g^{\prime 2}\right]\phi (x)=E\phi(x)~.
\label{phi}
\end{eqnarray}
For physical systems which we are interested in, the phase factor
$\exp(-g(x))$ is responsible for the asymptotic behaviors of the
wave function so as to ensure normalizability. The function
$\phi(x)$ satisfies a Schr\"odinger equation with a gauge
transformed Hamiltonian
\begin{eqnarray}
H_G=-\frac{d^2}{dx^2} + 2W_0(x)\frac{d}{dx}  +\left[V(x)
+W_0^\prime - W_0^2\right]~, \label{HG}
\end{eqnarray}
where $W_0(x)=g^\prime (x)$.  Now if $V(x)$ is such that the
quantal system is QES, that means the gauge transformed
Hamiltonian $H_G$ can be written as a quadratic combination of the
generators $J^a$ of some Lie algebra with a finite dimensional
representation.  Within this finite dimensional Hilbert space the
Hamiltonian $H_G$ can be diagonalized, and therefore a finite
number of eigenstates are solvable.  For one-dimensional QES
systems the most general Lie algebra is $sl(2)$
(\cite{TU}-\cite{Shifman}).  Hence if eq.(\ref{HG}) is QES then it
can be expressed as
\begin{eqnarray}
H_G=\sum C_{ab}J^a J^b + \sum C_a J^a + {\rm constant}~,
\label{H-g}
\end{eqnarray}
where $C_{ab},~C_a$ are constant coefficients, and the $J^a$ are
the generators of the Lie algebra $sl(2)$ given by
\begin{eqnarray}
J^+ &=& z^2 \frac{d}{dz} - Nz~,\cr
J^0&=&z\frac{d}{dz}-\frac{N}{2}~,~~~~~~~~N=0,1,2\ldots\cr J^-&=&
\frac{d}{dz}~.
\end{eqnarray}
Here the variables $x$ and $z$ are related by $z=h(x)$, where
$h(\cdot)$ is some (explicit or implicit) function . The value
$j=N/2$ is called the weight of the differential representation of
$sl(2)$ algebra, and $N$ is the degree of the eigenfunctions,
which are polynomials in a $(N+1)$-dimensional Hilbert space.

The requirement in eq.(\ref{H-g}) fixes $V(x)$ and $W_0(x)$, and
$H_G$ will have an algebraic sector with $N+1$ eigenvalues and
eigenfunctions. In this sector the eigenfunction has the general
form
\begin{eqnarray}
\psi=(z-z_1)(z-z_2)\cdots (z-z_N)\exp(-\int^z W_0(x) dx)~,
\label{psi-1}
\end{eqnarray}
where $z_i$ ($i=1,2,\ldots,N$) are $N$ parameters that can be
determined from the eigenvalue equations, namely, the Bethe ansatz
equations corresponding to the QES problem \cite{Ush,Ho}.  One can
rewrite eq.(\ref{psi-1}) as
\begin{eqnarray}
\psi =\exp(-\int^z W_N(x,\{z_i\}) dx)~,
\end{eqnarray}
and
\begin{eqnarray}
W_N(x,\{z_i\}) =
W_0(x) -  \sum_{i=1}^N \frac{h^\prime(x)}{h(x)-z_i}~. \label{W}
\end{eqnarray}
There are $N+1$ possible functions $W_N (x,\{z_i\})$ for the $N+1$
sets of eigenfunctions $\psi$. It is easy to check that $W_N$
satisfies the Ricatti equation \cite{Shifman,Roy}
\begin{eqnarray}
W^2 - W^\prime = V - E~, \label{Ricatti}
\end{eqnarray}
where $E$ is the eigenenergy corresponding to the eigenfunction
$\psi$ given in eq.(\ref{psi-1}) for a particular set of
parameters $\{z_i\}$.  Eq.(\ref{Ricatti}) shows the connection
between SUSY and QES problems.

From eqs.(\ref{SE}), (\ref{V}) and (\ref{Ricatti}) it is clear how
one should proceed to determine the magnetic fields so that the
Pauli equation becomes QES based on $sl(2)$: one needs only to
determine the superpotentials $W(x)$ according to
eq.(\ref{Ricatti}) from the QES potentials $V(x)$  classified by
Turbiner \cite{Tur2}. This is easily done by observing that the
superpotential $W_0$ corresponding to $N=0$ is related to the
gauge function $g(x)$ associated with a particular class of QES
potential $V(x)$ by $g^\prime (x)=W_0 (x)$. Once $W_0$ is
obtained, then $B_0=-W^\prime_0 (x)$ is the required magnetic
field that allows the weight zero ($j=N=0$) state to be known in
that class. But this state is just the ground state, and hence we
have not gone beyond the result of \cite{AC} . What is more
interesting is to obtain higher weight states (i.e. $j>0$), which
will include excited states.  For weight $j$ ($N=2j$) states, this
is achieved by forming the superpotential $W_N(x,\{z_i\})$
according to eq.(\ref{W}). Of the $N+1$ possible sets of solutions
of the Bethe ansatz equations, the set of roots
$\{z_1,z_2,\ldots,z_N\}$  to be used in eq.(\ref{W}) is chosen to
be the set for which the energy of the corresponding state is the
lowest (usually it is the ground state).  The required magnetic
field which gives rise to the $N+1$ solvable states is then
obtained as $B_N=-W_N^\prime$. From the table in \cite{Tur2} it is
easily seen that only six classes need be considered, namely class
I to class VI.  Class VII to IX are excluded because these are QES
systems with basic variables defined only on the half-line
$(0,\infty)$, while class X corresponds to periodic potentials
giving rise to non-normalizable wave functions. Below we shall
illustrate our construction of QES magnetic fields through the
class I and II QES systems, which serve as representative examples
of two different types of QES problems.

\vskip 1.0 cm \centerline{\bf Class I}

According to Turbiner's classification, the QES potential
belonging to class I has the form \cite{note}
\begin{eqnarray}
V_N(x)=a^2 e^{-2\alpha x}-
a\left[\alpha(2N+1)+2b_k\right]e^{-\alpha x}+
c\left(2b_k-\alpha\right)e^{\alpha x} + c^2 e^{2\alpha
x}+b_k^2-2ac~. \label{V-I}
\end{eqnarray}
Here $b_k\equiv b+k$ with constant $b$. Without loss of
generality, we assume $\alpha, a,c>0$ for definiteness.  The
corresponding gauge function $g(x)$ is given by
\begin{eqnarray}
g(x)= \frac{a}{\alpha} e^{-\alpha x} + \frac{c}{\alpha} e^{\alpha
x} + b_k x~. \label{g-I}
\end{eqnarray}
One should always keep in mind that the parameters selected must
ensure convergence of the function $\exp(-g(x))$ in order to
guarantee normalizability of the wave function (this is generally
not required by the mathematicians). We have also added the
constant $(b_k^2-2ac)$ in $V_N$ so that for $j=0$, the energy of
the ground state is zero ($E=0$). This is not necessary, but it
allows the results for $j=0$ and $j>0$ be presented in a unified
way. The potential $V(x)$ that gives the ground state is generated
by
\begin{eqnarray}
V(x)&=&V_0 -E\cr &=& W_0^2-W_0^\prime~,
\end{eqnarray}
with
\begin{eqnarray}
 W_0(x)&=&g^\prime(x)\cr &=&-a e^{-\alpha x}+ c e^{\alpha
x}+b+k~.
\end{eqnarray}
The corresponding magnetic field is given by
\begin{eqnarray}
B_0&=&-W_0^\prime (x)\cr & =& -a\alpha e^{-\alpha x} - c\alpha
e^{\alpha x}~. \label{B-I}
\end{eqnarray}

To obtain magnetic fields and the corresponding potentials which
admit solvable states with higher weights $j$, we must first
derive the Bethe ansatz equations. To this end, let us perform the
change of variable $z=h(x)=\exp(-\alpha x)$. Eq.(\ref{phi}) then
becomes
\begin{eqnarray}
-\alpha z^2 \frac{d^2\phi(z)}{dz^2} +\left[2az^2 -(2b_k+\alpha)z
-2c\right]\frac{d\phi(z)}{dz} +\left[-2aNz -
\frac{E}{\alpha}\right]\phi(z) =0~, \label{phi-I}
\end{eqnarray}
which can be written as a quadratic combination of the $sl(2)$
generators $J^+,~J^-$ and $J^0$ as
\begin{eqnarray}
T_I\phi&=&0~,\cr T_I&=&-\alpha J^+ J^- + 2a J^+ -
\left[\alpha(N+1)+2b_k\right]J^0 -2cJ^- + \rm{constant}~.
\label{T-I}
\end{eqnarray}

For $N>0$, there are $N+1$ solutions which include excited states.
Assuming $\phi(z)=\prod_{i=1}^N (z-z_i)$ in eq.(\ref{phi-I}), one
obtains the Bethe ansatz equations which determine the roots
$z_i$'s
\begin{eqnarray}
2az_i^2 -(2b_k+\alpha)z_i - 2c -2\alpha \sum_{l\neq
i}\frac{z_i^2}{z_i-z_l} =0~, \quad\quad i=1,\ldots,N~,
\label{BA-I}
\end{eqnarray}
and the equation which gives the energy in terms of the roots
$z_i$'s
\begin{eqnarray}
E =2\alpha c\sum_{i=1}^N \frac{1}{z_i} ~. \label{E-I}
\end{eqnarray}
Each set of \{$z_i$\} determine a QES energy $E$ with the
corresponding polynomial $\phi$.

As an example, consider the $j=1/2$ case with $N=1$ and
$\phi(z)=z-z_1$.  There are two solutions.  From eq.(\ref{BA-I}),
one sees that the root $z_1$ satisfies
\begin{eqnarray}
2az_1^2 -(2b_k+\alpha)z_1 -2c =0~,
\end{eqnarray}
which gives two solutions
\begin{eqnarray}
z_1^\pm=\frac{(2b_k+\alpha)\pm\sqrt{(2b_k+\alpha)^2 + 16ac}}{4a}~.
\end{eqnarray}
The corresponding energy is
\begin{eqnarray}
E^\pm&=&  2\alpha c\frac{1}{z_1}\cr &=&
-\frac{\alpha}{2}\left[(2b_k+\alpha)\mp\sqrt{(2b_k+\alpha)^2 +
16ac}\right]~.
\end{eqnarray}
For the parameters assumed here, the solution with root
$z_1^-=-|z_1^-|<0$ gives the ground state, while that with root
$z_1^+>0$ gives the first excited state. The superpotential $W_1$
is constructed according to eq.(\ref{W})
\begin{eqnarray}
W_1(x)&=&W_0-\frac{h^\prime(x)}{h(x)-z_1^-}\cr &=&-ae^{-\alpha x}+
c e^{\alpha x}+\frac{\alpha}{1+|z_1^-|e^{\alpha x}}+b+k~.
\end{eqnarray}
This gives the magnetic field
\begin{eqnarray}
B_1&=&-W_1^\prime (x)\cr & =& -a\alpha e^{-\alpha x} - c\alpha
e^{\alpha x}+\alpha^2\frac{|z_1^-|e^{\alpha x}}
{(1+|z_1^-|e^{\alpha x})^2}~,
\end{eqnarray}
and the potential that allows these two solvable states is
\begin{eqnarray}
V(x) &=& W_1^2-W_1^\prime~,\cr &=&V_1 -E^-~.
\end{eqnarray}
With this potential, the ground state and the excited state have
energy $E=0$ and $E=E^+ -E^-=\alpha \sqrt{(2b_k+\alpha)^2 +
16ac}$, respectively.  Our construction, based on the connection
between SUSY and QES systems, always sets the energy of the lowest
energy state to zero.

This example should convey the general ideas of our construction.
QES potentials and magnetic fields for higher degree $N$ are
obtained in the same manner. We note that even for higher values
of $N$ the equation (\ref{BA-I}) still remains an algebraic
equation whose solutions can always be found albeit may be
numerically. But even then the system remains a QES one.

We mention here that QES systems belonging to Class IV and VI can be
considered in a similar manner.

\vskip 0.5cm
\centerline{\bf Class II}

We shall consider one more class of QES potential, namely, class
II of Turbiner's classification.  The general procedure is the
same as that applied to class I.  But unlike class I, IV and
VI, which are called the first type QES problems, class II, III
and V belong to the second type.  In the first type QES problems,
$N+1$ eigenstates are solvable for a fixed potential with a fixed
degree $N$. For the second type, on the other hand, there are
$N+1$ QES potentials differing by the values of parameters and
have the same eigenvalue of the $i$-th eigenstate in the $i$-th
potential. For our present problem, each potential corresponds to
a magnetic field. Below we shall demonstrate this using class II
potentials.

The general form of class II QES potential is
\begin{eqnarray}
V_N(x,\lambda)=d^2 e^{-4\alpha x}&+& 2ad e^{-3\alpha x}+
\left[a^2+ 2d\left(b_k -\alpha(N+1)\right)\right]e^{-2\alpha x}\cr
&+&\left(2ab_k -\alpha a +\lambda \right)e^{-\alpha x}+b_k^2~.
\label{V-II}
\end{eqnarray}
The gauge function is
\begin{eqnarray}
g(x)= \frac{d}{2\alpha} e^{-2\alpha x} + \frac{a}{\alpha}
e^{-\alpha x} - b_k x~. \label{g-II}
\end{eqnarray}
As mentioned before, the parameters must be so chosen as to
guarantee the normalizability of the wave function.  For
definiteness, we assume $\alpha, d>0$ and $b_k=b+k <0$.

Letting $z=h(x)=\exp(-\alpha x)$, the equation of $\phi(z)$ is
\begin{eqnarray}
-\alpha z \frac{d^2\phi(z)}{dz^2} +\left[2dz^2
+2az+2b_k-\alpha\right]\frac{d\phi(z)}{dz} +\left[-2dNz
+\frac{\lambda}{\alpha} - \frac{E}{z\alpha}\right]\phi(z) =0~.
\label{phi-II}
\end{eqnarray}
The differential operator in eq.(\ref{phi-II}) can also be written
as
\begin{eqnarray}
T_{II}=-\alpha J^0 J^- + 2d J^+ + 2aJ^0 -
\left[\alpha\left(\frac{N}{2}+1\right)-2b_k\right]J^- +
\rm{constant}~. \label{T-II}
\end{eqnarray}
The energy $E$ and the parameter $\lambda$ are given by
\begin{eqnarray}
E&=&0~,\cr \lambda &=& \alpha \left(2b_k
-\alpha\right)\sum_{i=1}^N \frac{1}{z_i}~, \label{E-II}
\end{eqnarray}
where the $z_i$'s are to be solved from the Bethe ansatz equations
\begin{eqnarray}
dz_i^2+az_i + b_k -\frac{\alpha}{2} -\alpha\sum_{l\neq
i}\frac{z_i}{z_i-z_l} =0~, \quad\quad i=1,\ldots,N~. \label{BA-II}
\end{eqnarray}
The required magnetic field is again given by the roots $z_k$'s
through eq.(\ref{W}).  For $N=0$, one has $\lambda=0$.

So far everything appears to be the same as for class I.  The main
point to note is that $V_N(x,\lambda)$ is a function of the
parameter $\lambda$ as well as $N$, and $\lambda$ is determined
from the roots $\{z_1,z_2,\ldots,z_N\}$.  For a fixed $N$ there
are $N+1$ possible sets of the roots.  Therefore one can construct
$N+1$ possible potentials $V_N^{(m)} (x)$ ($m=0,1,\ldots,N$) for
eq.(\ref{SE})  according to
\begin{eqnarray}
V_N^{(m)} &=& (W^{(m)}_N)^2 - (W^{(m)}_N)^\prime\cr &=&
V_N(x,\lambda^{(m)})- E~, ~~~~~~~~~~E=0~.
\label{V-m}
\end{eqnarray}
Here $\lambda^{(m)}$ is the parameter evaluated using the $m$-th
set of roots of the Bethe ansatz equations in eq.(\ref{E-II}), and
$W^{(m)}_N$ is obtained from eq.(\ref{W}) using the same set of
roots.  We recall here that in class I discussed previously, the
superpotential $W_N$ was calculated using the set of roots which
gives the lowest energy, but here all the $N+1$ sets of roots have
to be used. For each potential $V_N^{(m)}$ only one state is solved
(by the $m$-th set of roots) with the same energy $E=0$. And each
potential $V^{(m)}$ corresponds to a magnetic field
$B^{(m)}_N=-(W^{(m)}_N)^\prime$. For the family of potentials,
only some (generally one) potentials have ground state solved,
while for others the solvable state is an excited state. In other words,
we
have a family of magnetic fields for which Pauli equation is solvable
for one level with the same energy.

To illustrate these points, let us now give two examples.
Consider first the case for $N=0$, giving only the ground state
$\psi(x)=const.
\times \exp(-g(x)$ with $g(x)$ being given by eq.(\ref{g-II}). The
energy of
this state is $E=0$.  In this case the parameter $\lambda$ is
$\lambda=0$, and the QES potential that gives rise to this solvable
ground
state is
\begin{eqnarray}
V(x)=V_0^{(0)}=V_0(x, \lambda=0)~,
\end{eqnarray}
which according to eq.(\ref{V-m}) is generated by
the superpotential
\begin{eqnarray}
W_0^{(0)} (x) &=& g^\prime (x)\nonumber\cr &=& -d e^{-2\alpha x}
- a e^{-\alpha x}-b-k~.
\end{eqnarray}
The corresponding magnetic field is
\begin{eqnarray}
B_0&=&-W_0^{(0)\prime} (x)\cr & =& -2d\alpha e^{-2\alpha x} -
a\alpha e^{-\alpha x}~.
\end{eqnarray}

Now we come to the case for $N=1$.  The QES wave function is
$\psi= (z-z_1)\exp(-g)$, and the energy of the state is $E=0$. The
root $z_1$ is solved from the Bethe ansatz equation (\ref{BA-II})
\begin{eqnarray}
dz_1^2+az_1 + b_k -\frac{\alpha}{2}  =0~. \label{BA-N1}
\end{eqnarray}
We recall here that we have assumed $\alpha, d>0$ and $b_k=b+k
<0$.  Eq.(\ref{BA-N1}) gives two solutions
\begin{eqnarray}
z_1^{(0,1)}=\frac{-a\mp\sqrt{a^2 + 4 d (|b_k|+\alpha/2)}}{2d}~,
\end{eqnarray}
where $z_1^{(0)}$ ($z_1^{(1)}$) corresponds to the solution given
by the minus (plus) sign.  Since $z_1^{(0)} <0$, the state $\psi$
constructed with $z_1^{(0)}$ is the ground state, while that with
$z_1^{(1)}$ is the first excited state.   According to
eq.(\ref{V-m}), the ground state is the only QES state for the
system with potential $V_1^{(0)}=V_1(x, \lambda^{(0)}) $ and the
excited state is the only QES state for the potential
$V_1^{(1)}=V_1(x, \lambda^{(1)})$, where the parameters $\lambda
^{(0,1)}$ are given by
\begin{eqnarray}
\lambda^{(0,1)} = -\frac{\alpha \left(2|b_k|
+\alpha\right)}{z_1^{(0,1)}} ~.
\end{eqnarray}
These two potentials are generated by the superpotentials
\begin{eqnarray}
W_1^{(m)} (x)&=&W_0^{(0)}-\frac{h^\prime(x)}{h(x)-z_1^{(m)}}\cr
&=&-de^{-2\alpha  x}- a e^{-\alpha x}+\frac{\alpha}{1-z_1^{(m)}
e^{\alpha x}}-b-k~,~~~~m=0,1~,
\end{eqnarray}
with the corresponding magnetic fields being
\begin{eqnarray}
B_1^{(m)}&=&-W_1^{(m)\prime} (x)\cr & =& -2d\alpha e^{-2\alpha x}
- a\alpha e^{-\alpha x}-\alpha^2\frac{z_1^{(m)}e^{\alpha x}}
{(1-z_1^{(m)}e^{\alpha x})^2}~,~~~~m=0,1~.
\end{eqnarray}
The point to note is that the energy of the ground state for the
potential $V_1^{(0)}$ and that of the first excited for
$V_1^{(1)}$ are both equal to zero, i.e. $E=0$.

The case for class III and V are the same as the present one.
We shall not repeat the arguments here.

\section{Magnetic field in symmetric gauge}

We now consider the same problem in the symmetric gauge
\begin{eqnarray}
A_x= yf(r)~,~~ A_y=-xf(r)~,
\end{eqnarray}
where $r^2=x^2+y^2$.  The magnetic field $B_z=B$ is then given by
\begin{eqnarray}
B(x,y)= -2f(r)- rf^\prime(r)~. \label{B2}
\end{eqnarray}
 The Pauli Hamiltonian is
\begin{eqnarray}
H=-\left(\frac{d^2}{dx^2}+\frac{d^2}{dy^2}\right) +r^2f^2 -2fL_z
-\left(2f+rf^\prime\right)\sigma_z~. \label{H2}
\end{eqnarray}
Here $L_z$ is the $z$-component of the orbital angular momentum,
and $f^\prime=df/dr$. We assume the wave functions to have the
form
\begin{eqnarray}
 \Psi(t,{\bf x}) = \frac{1}{\sqrt{r}}
\psi_m(r, \varphi)
\end{eqnarray}
with
\begin{eqnarray}
\psi_m(r, \varphi) = \left( \begin{array}{c} R_1(r)e^{im\varphi}\\
R_2(r)e^{i(m+1)\varphi}
\end{array}\right)
\label{wf}
\end{eqnarray}
with integral number $m$. The function $\psi_m(r,\varphi)$ is an
eigenfunction of the conserved total angular momentum $J_z=L_z +
\sigma_z/2$ with eigenvalue $J=m+1/2$.  The components $R_1$ and
$R_2$ satisfy
\begin{eqnarray}
\left[-\frac{d^2}{dr^2}+r^2f^2 -2f\left(m+1\right)-rf^\prime +
\frac{m^2-\frac{1}{4}}{r^2}\right] R_1(r) = ER_1(r)~,\label{R1}
\end{eqnarray}
and
\begin{eqnarray}
 \left[-\frac{d^2}{dr^2}+r^2f^2 -2fm + rf^\prime +
\frac{(m+1)^2-\frac{1}{4}}{r^2}\right] R_2(r)= ER_2(r)~,\label{R2}
\end{eqnarray}
where $E$ is the energy.

The gauge function $g(r)$ for eq.(\ref{R1}), which accounts for
the asymptotic behaviors of the system, has the general form
\begin{eqnarray}
g(r)=\int^r \rho f(\rho) d\rho -\gamma\ln
{r}~,~~~~~\gamma=|m|+1/2~. \label{g}
\end{eqnarray}
The corresponding superpotential $W(r)$ is
\begin{eqnarray}
W(r)&=&g^\prime (r)\cr &=& rf(r)-\frac{\gamma}{r}~.\label{W2}
\end{eqnarray}
One can check that the potentials in eq.(\ref{R1}) and (\ref{R2})
are given by $V_-$ and $V_+$, respectively, where $V_\mp=W^2\mp
W^\prime$ for positive $m\geq 0$.  This again shows the SUSY
structure of the Pauli equation.  Hence, the procedure presented
in the last section can also be applied in this case (for $m\geq
0$).

Our task is to find the form of $f(r)$ such that the Pauli
equation is QES.  It is seen that both eq.(\ref{R1}) and
(\ref{R2}) are in the Schr\"odinger form. So one could try to find
$f(r)$ that would fit eq.(\ref{R1}) and (\ref{R2}) into the forms
classified in \cite{Tur2}. As before, we shall only consider the
upper component $R_1$. The lower component can be obtained by
SUSY.  We found that there exist three forms of magnetic fields
which make the Pauli equation QES.  These three forms of magnetic
fields give QES potentials that belong to class VII, VIII and IX
of Turbiner's classification. Below we shall discuss only the case
for class VII.   The other two classes can be considered in a similar
manner.

\vskip 0.5cm
\centerline{\bf Class VII}

By inspection one finds that if we let $f(r)=f_0(r)=ar^2+b$
($a>0,b$ are constants), then eq.(\ref{R1}) belongs to class VII
of Turbiner's classification with $N=0$. With this form of the
function $f$, the potential in eq.(\ref{R1}) is
\begin{eqnarray}
V(r)=a^2r^6+2abr^4+\left[b^2-2a\left(m+2\right)\right]r^2
+\gamma\left(\gamma-1\right)r^{-2}-2b\left(m+1\right)~. \label{V1}
\end{eqnarray}
The magnetic field is $B_0=-4ar^2-2b$.  The general potential in
class VII has the form
\begin{eqnarray}
V_N(r)=a^2r^6+2abr^4+\left[b^2-a\left(4N+2\gamma
+3\right)\right]r^2
+\gamma\left(\gamma-1\right)r^{-2}-b\left(2\gamma+1\right)~.
\label{cVII}
\end{eqnarray}
 Comparing eqs.(\ref{V1}) and (\ref{cVII}) one
concludes the potential (\ref{V1}) does belong to class VII with
$N=0$ and for $m\geq 0$.

As in the asymmetric case, we assume $R_1=\exp(-g(r))\phi$, then
$\phi$ satisfies the same equation (\ref{phi}) with all the
derivatives now being with respect to the variable $r$ instead of
$x$.  With the choice
\begin{eqnarray}
g(r)=\frac{a}{4}r^4 + \frac{b}{2}r^2 -\gamma\ln {r}~, \label{g4}
\end{eqnarray}
one can check that $V(r)$ in eq.(\ref{V1}) is generated by
$W_0(r)\equiv g^\prime (r)$ in the form $V=W_0^2 -W_0^\prime$.
Hence the method used in the asymmetric gauge can also be applied
here to generate magnetic fields which allow for QES potentials
with higher weight.  To proceed, we need to obtain the Bethe ansatz
equations for $\phi$.

Letting $z=h(r)=r^2$, eq.(\ref{phi}) becomes
\begin{eqnarray}
-4 z \frac{d^2\phi(z)}{dz^2} +\left[4az^2 +4bz -2\left(2\gamma
+1\right) \right]\frac{d\phi(z)}{dz} -\left[4aNz + E
\right]\phi(z) =0~. \label{phi-VII}
\end{eqnarray}
In terms of the $sl(2)$ generators $J^+,~J^-$ and $J^0$, the
differential operator in eq.(\ref{phi-VII}) can be written as
\begin{eqnarray}
T_{VII}=-4 J^0 J^- + 4a J^+ +4bJ^0 - 2\left(N+2\gamma +1\right)J^-
+ \rm{constant}~. \label{T-VII}
\end{eqnarray}
For $N=0$, the energy is $E=0$.  For higher $N>0$ and
$\phi(r)=\prod_{i=1}^N (z-z_i)$, the function $f(r)=f_N(r)$ is
obtained from eqs.(\ref{W}) and (\ref{W2}):
\begin{eqnarray}
f_N(r) = f_0(r) - \frac{1}{r} \sum_{i=1}^N
\frac{h^\prime(r)}{h(r)-z_i}~.
\end{eqnarray}
For the present case, the roots $z_i$'s are found from the Bethe
Ansatz equations
\begin{eqnarray}
2az_i^2 +2bz_i -\left(2\gamma+1\right) - \sum_{l\neq
i}\frac{z_i}{z_i-z_l} =0~, \quad\quad i=1,\ldots,N~,
\label{BA-VII}
\end{eqnarray}
and the energy in terms of the roots $z_i$'s is
\begin{eqnarray}
E=2\left(2\gamma+1\right)\sum_{i=1}^N \frac{1}{z_i}~.
\label{E-VII}
\end{eqnarray}

For $N=1$ the roots $z_1$ are
\begin{eqnarray}
z_1^\pm=\frac{-b\pm\sqrt{b^2+2a(2\gamma+1)}}{2a}~. \label{z1}
\end{eqnarray}
The energies are
\begin{eqnarray}
E^\pm=2(b\pm\sqrt{b^2+2a(2\gamma+1)})~.
\end{eqnarray}
 For $a>0$, the root $z_1^-=-|z_1^-|<0$ gives the
ground state. With this root, one gets the superpotential
\begin{eqnarray}
W_1(r)=ar^3 +br -\frac{2r}{r^2+|z_1^-|} -\frac{\gamma}{r}~.
\end{eqnarray}
The QES potential appropriate for the Pauli problem is
\begin{eqnarray}
V(x) &=& W_1^2-W_1^\prime~,\cr &=&V_1 -E^-~.
\end{eqnarray}
With this choice of the potential, the ground state and the
excited state have energy $E=0$ and $E=E^+
-E^-=4\sqrt{b^2+2a(2\gamma+1)}$.  The magnetic field $B_1$ is
calculated from eq.(\ref{B2}) using the function
\begin{eqnarray}
f_1(r)&=& \frac{1}{r}\left[W_1(r)+\frac{\gamma}{r}\right]\cr
 &=&ar^2
+b -\frac{2}{r^2+|z_1^-|}~,
\end{eqnarray}
which gives
\begin{eqnarray}
B_1(r)=-4ar^2-2b+ \frac{4|z_1^-|}{(r^2+|z_1^-|)^2}~.
\end{eqnarray}

Just as class I, class VII is also of the first type.  On the
other hand, class VIII and IX belong to the second type.  We will
not repeat the discussions here.

\section{\bf Summary and Discussions}
In this paper an attempt to give a QES generalization of the
result of Aharonov  and Casher is presented.  We have given a
general procedure for determining  possible (nonuniform) magnetic
fields such that the Pauli equation becomes  QES based on the
$sl(2)$ algebra.  This procedure makes full use of the close
connection between QES systems and SUSY.  Of the ten classes of
$sl(2)$-based  one-dimensional QES systems, we have found that
only  nine classes allow such construction. It would be interesting
to extend our procedure to the Dirac equation.

The Pauli equation is supersymmetric owing to the fact that the
gyromagnetic ratio is two, i.e. $g=2$.  We would like to mention
that recently it was realized \cite{KP1,KP2} that if one changes
$g$ to some unphysical values $g=2n$ ($n$ positive integers), then
for magnetic field of special exponential  and quadratic forms,
the generalized Pauli equation could possess a new type of
supersymmetry \cite{N-SUSY,Aoyama,KP3}.  This is the nonlinear
generalization of the usual supersymmetry, and is given the name
``nonlinear  holomorphic supersymmetry" in \cite{KP1,KP2,KP3}, or
`` $n$-fold supersymmetry" in \cite{Aoyama}.  It is characterized
by a non-linear superalgebra among the supercharges and the
Hamiltonian, and the anticommutator of the supercharges is a
polynomial of the Hamiltonian. The usual SUSY can be viewed as a
special case, namely, the $n=1$ case of the $n$-fold SUSY. Soon
after its discovery, the $n$-fold SUSY was shown to be closely
related to quasi-exact solvability \cite{Aoyama,KP3}. For the
generalized Pauli equations considered in \cite{KP1,KP2}, the
weight $j=N/2$ characterizing the quasi-exact solvability of the
system is given in terms of the number $n$ of the $n$-fold SUSY
and some parameters of the system. The authors of \cite{KP2} found
certain duality transformations which mix the number $n$ and the
parameters to give different values of $N$. These duality
transformations thus connect different sectors of the generalized
Pauli equations.  From a mathematical point of view, quasi-exact
solvability of the generalized Pauli equation is an interesting
subject to be further explored.  It is worth mentioning that the
main difference between the generalized Pauli equation on a plane
considered in \cite{KP1} and the system considered by us is that
in \cite{KP1} the weight $j=N/2$ is related to the $n$-fold SUSY
by $n=2j+1=N+1$, while in our case $n$ is always one (i.e. $n=1$)
and $N$ can be chosen arbitrarily ($N=0,1,2,\ldots$).  Hence when
the system in \cite{KP1} is reduced to our case (by setting
$n=1$), the only QES state that is retained is the ground state
(corresponding to $N=0$). Furthermore, since the number $N$ in our
case is an arbitrarily chosen number, the kind of duality
transformation obtained in \cite{KP2} does not exist in our
system.

Finally we mention a few things about the degeneracy of the energy
levels.
First we note that the Hamiltonians $H_{\pm}$ are SUSY partners (since
they
are built from {\it nodeless} superpotentials) and thus $H_+$ shares all
the
levels of $H_-$ except the zero energy state. Therefore all the levels
of
$H_-$ are doubly degenerate except the zero energy level. This is in
agreement
with the results of ref \cite{AC}.
We now come to the question of degeneracy of the levels within one
component,
namely, $H_-$. Since in all the cases considered in this paper the
magnetic
fields are nonuniform, so according to \cite{grosse} the excited states
are
nondegenerate. This behaviour of the excited states is in contrast to
the
ground state which is always degenerate with the degeneracy depending on
the
magnetic flux.

\vskip 2cm \centerline{\bf Acknowledgments}

This work was supported in part by the Republic of China through
Grant No. NSC 91-2112-M-032-010.  We would like to thank Professor
Turbiner for posing the question whether one could give a QES
generalization of the result of Aharonov and Casher, which
eventually resulted in this work. P.R. would like to thank the
Department of Physics at Tamkang University for support during
his visit.  C.L.H. would also like to thank Professor Y. Hosotani
for general discussions on the subject, and Dr. T.Tanaka for
discussions on $\cal N$-fold supersymmetry.  Finally the authors would
like to thank the referees for their constructive criticism.


\begin{thebibliography}{2}
\bibitem{note}  We follow the notations used in
\cite{Tur2} as closely as possible for the six classes discussed
in this paper.  We also take this opportunity to correct some
typographical errors in that paper.
\end{thebibliography}
\end{document}